\RequirePackage{fix-cm}
\documentclass[smallcondensed]{svjour3}     % onecolumn (ditto)
\smartqed  % flush right qed marks, e.g. at end of proof
\usepackage{graphicx}
\usepackage{amsmath}
\usepackage[outline]{contour}
\usepackage[misc]{ifsym}

\begin{document}

\title{Preference Modeling by Exploiting Latent Components of Ratings%\thanks{Grants or other notes
}


\author{Junhua Chen \and
        Wei Zeng   \and \\
        Junming Shao    \and
        Ge Fan
}


\institute{Junhua Chen \and Wei Zeng (\Letter) \and Ge Fan
            \at
              Trusted Computing and Automated Reasoning Lab, University of Electronic Science and Technology of China, Chengdu 611731, P. R. China \\
              \email{chenjunhua@std.uestc.edu.cn;  zwei504@gmail.com; fange@std.uestc.edu.cn}%  \\
           \and
           Junming Shao \at
              Web Sciences Center, University of Electronic Science and Technology of China, Chengdu 611731, P. R. China \\
              \email{junmshao@uestc.edu.cn}
           }

\date{Received: date / Accepted: date}

\maketitle

\begin{abstract}
Understanding user preference is essential to the optimization of recommender systems. As a feedback of user's taste, rating scores can directly reflect the preference of a given user to a given product. Uncovering the latent components of user ratings is thus of significant importance for learning user interests. In this paper, a new recommendation approach, called LCR, was proposed by investigating the latent components of user ratings. The basic idea is to decompose an existing rating into several components via a cost-sensitive learning strategy. Specifically, each rating is assigned to several latent factor models and each model is updated according to its predictive errors. Afterwards, these accumulated predictive errors of models are utilized to decompose a rating into several components, each of which is treated as an independent part to retrain the latent factor models. Finally, all latent factor models are combined linearly to estimate predictive ratings for users. In contrast to existing methods, LCR provides an intuitive preference modeling strategy via multiple component analysis at an individual perspective. Meanwhile, it is verified by the experimental results on several benchmark datasets that the proposed method is superior to the state-of-art methods in terms of recommendation accuracy.
\end{abstract}

% Section 1
\section{Introduction}
\label{Intro}
With the rapid growth of the Internet and the overwhelming amount of contents and choices that people are confronted with, recommender systems have been developed to facilitate the decision making process. During the past decades, more and more researchers have started to study the multi-criteria recommender system, which allows individual users to rate multiple attributes of an item. For instance, a two-criteria movie recommender system allows users to express their preferences on two attributes of a movie (e.g. story novelty and visual effect). A user may be fond of the visual effects, but dislike the story of a movie. In such a case, the movie is rated by a user based on two ratings, namely the story and visual effect attribute.

In a multi-criteria recommender system, an individual user can make a choice based on more than one utility-related aspects.  Actually, a user usually rates a movie after a comprehensive consideration. For example, she may firstly consider the movie's director, actors, story and visual effects, and then make the choice. Therefore, the accuracy of item recommendation can be  enhanced by the additional information provided by multi-criteria ratings because it can represent more complex preferences of each users. Recent works also demonstrated that the multi-criteria technique is superior to the traditional methods that utilize single-criteria ratings \cite{Adomavicius_Multi_Criteria_2007}.

However, nowadays, most existing websites don't adopt the mechanism of rating multiple attributes of an item. Most collected datasets accordingly only contain single-criteria ratings. Motivated by the technique of multi-criteria recommender system, we supposed that users' preferences are multi-dimensional and their ratings consist of multiple latent components even in a single-criteria recommender system. And each latent component is treated as an independent part which reflects one preference of a user. In order to uncover these latent components, a latent factor model based on the matrix factorization approach was adopted to map both users and items to a joint latent factor space. In the traditional matrix factorization approach, all ratings are trained by a single model. However, each rating is assigned to several latent factor models simultaneously in the training process in our method. And the weights of each latent factor model are computed according to their predictive errors. After the training process, those models possess varied weights which can represent the bias of users' preferences. Then, weights of models are used to decompose ratings.

In a traditional multi-criteria recommender system, all ratings on attributes of an item are explicitly given by individual users, whereas in our method, all latent components of a rating are automatically learned by several latent factor models. Therefore, our method not only overcome the drawback of single-criteria ratings but also does not require so much information as the multi-criteria technique. In a word, the contributions of our work can be summarized as follows:

\textbf{Intuitive preference model}. Instead of utilizing multiple ratings explicitly given by users, we decompose a rating into multiple latent components. Those latent components can better reflect the user's preference than a single-criteria rating. Moreover, less information is required by our method than that of the multi-criteria recommender system. Therefore, our method enjoys better application possibilities.

\textbf{High prediction performance}. Our original motivation is to improve the algorithm's accuracy with the user ratings data, by borrowing ideas from the multi-criteria technique. With our method, more information can be uncovered from individual ratings, which can solve the sparsity problem of recommender systems to some extent. Meanwhile, the experimental results showed that our method enjoys a great improvement over the state-of-the-art methods when fewer ratings are provided in the training datasets. Moreover, users' preferences can be more thoroughly uncovered by our method, which contributes to more accurate item recommendations for users.

\textbf{Efficiency}. Although our method brings in several latent models to learn the latent components of individual ratings, its computation time grows linearly with the single latent model. If the computation time of a single model is supposed to be $t$, then that of our method is $(c+1)t$, where $c$ is the number of latent models. Firstly, $t$ time is required to train the latent models. Secondly, the learned models in the previous step are retrained by using the latent components of ratings, which requires $ct$ time. As a matter of fact, the second training process can be executed distributively, which can further reduce the computation time of our method.

The rest of the paper is organized as follows. Section \ref{Related} gives a brief overview of the related research close to our work. Section \ref{LCR} covers preliminaries and our proposed model. The Experiments are shown in section \ref{Exp} and finally, this paper concludes with section \ref{Conclusion}.

% Section 2
\section{Related works}
\label{Related}

Motivated by the multi-criteria technique, we decomposed each single-criteria rating into several components \cite{Adomavicius_Multi_Criteria_2007,Breese_UAI_1998}. Therefore, our work is related with both the single-criteria and multi-criteria approaches. In this section, the related works are reviewed.

\subsection{Single-criteria recommender systems}
Up to now, plenty of methods have been proposed to solve the personalization problem. Among them, the collaborative filtering (CF) methods are considered to be the most popular one, which has been widely investigated and applied in online systems. Generally speaking, the CF methods can be categorized into two classes: neighborhood \cite{ICF2,1995_bellcore_CHI} and model-based methods \cite{Model_LSA,Model_LDA,Model_Boltzmann}. The neighborhood CF approaches can be further divided into two groups: the user-based CF and the item-based CF \cite{UCF1,UCF2,ICF1}. The assumption of the user-based CF is that similar users have similar tastes or preferences, whereas the item-based CF assumes that a user tends to collect similar items. The key of neighborhood CF approaches is to compute the similarity between neighbors or to provide a method to find the candidate neighbors. In order to find user's neighbors efficiently, some cluster-based methods have been proposed. Cluster-based methods first cluster the users and then calculate the user's neighbors within the cluster. For instance, Alqadah \cite{AlqadahRHA15} proposed a biclustering neighborhood CF approach for top-n recommendation which combines local similarity of biclusters with global similarity. Liu \cite{MC2} exploited the global k-means method to divide users into disjoint groups by making using of users' multi-criteria ratings.

Due to the simplicity and efficiency of neighborhood CF methods, they have a very wide application. For example, GroupLens and Bellcore video exploited individual ratings to predict their interests in articles and movies \cite{UCF1,UCF2}. Since the number of users are more than that of items in most systems and each item tends to have more ratings than each user. Item-based CF methods are preferable by online retailers, such as Amazon.com and Last.fm \cite{LastFM,ICF1,ICF2}.

Unlike neighborhood CF methods, model-based CF methods exploit users' ratings to learn a predictive model. Breese \cite{Breese_UAI_1998} proposed a Bayesian clustering model which introduces certain groups to reflect the common preferences and tastes of users. The latent semantic analysis models users' ratings as a mixtuer of individual communities \cite{Model_LSA}. LDA is a flexible generative probabilistic model for the collection of discrete data, which can be used for document classification and collaborative filtering \cite{Model_LDA}. The Restricted Boltzmann Machines, a kind of two-layer undirected graphical models, was presented to recommend movies for users \cite{Model_Boltzmann}.

It is worth mentioning that the matrix factorization (MF) approach is becoming more and more popular in recent years due to its good scalability and predictive accuracy \cite{Survey_MF,PMF,Bayesia_PMF,Low_rank,MC_MF,KannanIP14}. In the matrix factorization approaches, the rating matrix is composed of two parts, i.e. the user latent factor matrix and the item latent feature matrix. The predictive score is obtained by the computation based on inner product of the user latent feature vector and the item latent feature vector \cite{Survey_MF}. Usually, the low-rank approximation method and the regularization method are used to get the latent feature matrices and prevent overfitting, respectively \cite{PMF,Bayesia_PMF,Low_rank,MC_MF}. The probabilistic matrix factorization is a probabilistic interpretation of traditional matrix factorization methods \cite{PMF}, but the probabilistic approach can yield better scalability and robustness. A Bayesian treatment of probabilistic matrix factorization can yield a even more robust model compared to probabilistic matrix factorization by automatically controlling model capacity through the priors while still maintaining good scalability \cite{Bayesia_PMF}. Lee \cite{Low_rank} considered the locality in low-rank matrix factorization by assuming that the matrix is a representation of the observed matrix as a weighted sum of low-rank matrices. Following the locality assumption in matrix factorization, Wang \cite{MC_MF} proposed a multi-topic matrix factorization method that the locality in matrix is interpreted as topic. Kannan \cite{KannanIP14} proposed a bounded matrix factorization method which imposed a lower and an upper bound on every estimated missing element of the rating matrix. The matrix factorization is quite a flexible method, that it allows incorporation of additional information, such as time factor \cite{Time_factor1,Time_factor5,Time_factor6,train_test_split}, geographical information \cite{POI1,POI2,POI3,POI4} and social information \cite{Social1,Social2,Social3,Social4}. For instance, Koren \cite{Time_factor1} investigated the temporal dynamics of customer preferences and modeled the temporal dynamics along the whole time period. Authors applied the methodology with two recommender techniques: the factorization model and the neighborhood model. In both models, the temporal dynamics can be useful in improving the quality of rating predictions. McAuley \cite{train_test_split} developed a latent factor model which explicitly accounts for each user's level of experience. The time-aware model can not only achieve better recommendations but also allow to study the role of user experience and expertise. Lian \cite{POI4} incorporated the geographical information and the user activity data by the weighted matrix factorization which can alleviate the sparsity problem. Shen \cite{Social4} integrated the user-item ratings with the social information by a probabilistic model and the expectation–maximization algorithm was used to infer parameters of the model.

\subsection{Multi-criteria recommender systems}
Comparing to single-criteria recommender systems, multi-criteria recommender system contains more information, including ratings of item attributes. The complexity of algorithms is increased by the additional information, but in most cases, the quality of recommendation can also be improved by incorporating the auxiliary data \cite{Adomavicius_Multi_Criteria_2007,MC1,LeungCC06,MC2}. To the best of our knowledge, examples of multi-criteria recommender systems include Zagat's Guide, Buy.com and Yahoo! Movies \cite{MC_data2}. To exploit the information of multi-criteria ratings, a commonly used method is to extend the computation of similarity, from single-criteria ratings to multi-criteria ratings \cite{MC1,LeungCC06}. For example, within a user-based collaborative filtering approach, the similarity of two users is computed by making use of their single-criteria ratings while in a multi-criteria recommender system, the computation of similarity is performed on each criteria and finally average the result over all criteria. Lee \cite{MC_data1} extended the concept of single criteria rating to multi-criteria ones and utilized skyline queriy algorithm to find candidate items. Jannach \cite{MC1} made use of support vector regression to determine the relative importance of multi-criteria ratings and combined user-based and item-based regression model in a weighted way.

Another way to make use of multi-criteria ratings is to construct a predictive model by learning from the observed data, including probabilistic modeling, multi-linear SVD model and matrix factorization \cite{MC_MF}. For instance, Saboo \cite{MC_probalilistic_model} extended the flexible mixture model to multi-criteria rating systems, where the user behavior and item characteristics were characterized separately by two latent variables. Li \cite{MC_SVD} improved the traditional collaborative filtering method by expanding the criteria to a tensor and utilizing the multi-linear SVD model. McAuley \cite{mcauley2012learning} learned attitudes and attributes from explicit multi-aspect reviews with a joint probabilistic model to yield better recommendations.

% Section 3
\section{Latent components of ratings}
\label{LCR}

In this section, we mainly discuss our method, namely LCR model. Generally speaking, a rating system can be represented by a weighted adjancy matrix $R_{n\times m}=\{r_{ui}\}$, where $r_{ui}$ is the rating that user $u$ gives to item $i$, and $0$ otherwise. $n$ and $m$ are the number of users and the number of items in the system, respectively. As we know, the goal of a recommender system is to predict scores between each user and her uncollected items and then to recommend the top-$L$ items with the highest scores. Thus, research on recommender systems is always challenged by finding the most accurate recommendation algorithms.

\subsection{The problem statement}
In the multi-criteria recommender system, an individual user's diverse and complex preferences can be demonstrated by giving ratings to attributes of an item. However, in most existing recommender systems, a particular user is constrained to give a single-criterion value to an item. It is shown by recent works that this mechanism has potential limitations, because a user may make a choice based on more than one utility-related aspect. By borrowing ideas from multi-criteria recommender system, the goal of our work is to decompose the rating matrix $R$ into $c$ latent matrices: $R=\sum_{\alpha=1}^cR_{\alpha}$. We assumed that those latent matrices are independent to each other, and the latent matrix $R_{\alpha}$ has the same size with the original rating matrix $R$. There are a lot of decomposition forms. For instance, we can decompose $R$ randomly. However, this method is useless in improving the accuracy of recommendation algorithms. More details can be found in the \emph{Experiments} section.

In this paper, it is assumed that preferences of users are multi-dimensional, and each dimension can be represented by a latent factor model. It is just like, in a multi-criteria recommender system, different models are used to denote a user's different preferences on different attributes of an item. Thus, with our method, the complex preferences of users can be demonstrated with no necessity for additional information. However, most recommender systems do not have the mechanism for users to give ratings to attributes of items, and it is also inconvenient for users to do so. Moreover, to simplify the computation, those latent factor models are supposed to be independent of each other and are trained simultaneously. In our future work, the situation that latent models are dependent to each other will be considered. Then, potential approaches such as collective matrix factorization \cite{Singh_RLV_2008} and transfer learning \cite{Pan_STL_2010} can be taken into account.

In this paper, the cost sensitive approach was adopted to learn latent factor models. More specially, $c$ latent factor models,  $\{\Theta_1, \Theta_2, ..., \Theta_c \}$, were randomly initialized, and then the predictive score $\widehat{r}_{ui}^{(\alpha)}$ between user $u$ and item $i$ was computed by model $\Theta_{\alpha}$. By referring to the deviation between $\widehat{r}_{ui}^{(\alpha)}$ and the real rating $r_{ui}$, the weight of model $\Theta_{\alpha}$ with respect to rating $r_{ui}$ was computed. Finally, by exploiting the accumulated weight, the rating matrix $R$ was decomposed into $c$ latent matrices. All of the major notations used in this paper can be found in Table \ref{tab1:notations}.

\begin{table}[t]
{\def\arraystretch{1.6}
\begin{tabular}{ll}
\hline
Notation & Description   \\\hline \hline
$n$ & The number of users \\
$m$ & The number of items \\
$c$ & The number of latent component of ratings \\
$R$ & The rating matrix \\
$r_{ui}$ & The rating that user $u$ gives to item $i$ \\
$\widehat{r}_{ui}$ & The predictive rating between user $u$ and item $i$ \\
$x_u,y_i$ & The latent factor vector of user $u$ and item $i$, respectively \\
$\mu$ & The global average rating \\
$b_i$,$b_u$ & The deviation of user $u$ and item $i$, respectively \\
$\lambda$ & The coefficient of the regularization \\
$\gamma$ & The learning rate \\
$\widehat{r}_{ui}^{(\alpha)}$ & The predictive score between user $u$ and item $i$ by latent model $\Theta_{\alpha}$ \\
$x_u^{(\alpha)}$,$y_i^{(\alpha)}$ & The latent factor vector of user $u$ and item $i$ with respective to model $\Theta_{\alpha}$ \\
$w_{ui}^{(\alpha)}$ & The weight allocated for model $\Theta_{\alpha}$ given the training rating $r_{ui}$ \\
\hline
\end{tabular}}
\caption{Notations used in this paper.}
\label{tab1:notations}
\end{table}

\subsection{Latent factor models}
In this paper, latent factor models were adopted to decompose ratings, based on the matrix factorization. Firstly, users and items were mapped by the matrix factorization approach to a joint latent factor space of dimensionality $k$. Then each user was associated with a vector $x_u$, and each item was associated with a vector $y_i$. And the predictive score $\widehat{r}_{ui}$ between user $u$ and item $i$ was obtained by the inner product of $x_u$ and $y_i$:

\begin{equation}
\label{basic_mf}
\widehat{r}_{ui}=x_u^T y_i,
\end{equation}
which is the basic form of the matrix factorization. One advantage of the matrix factorization approach is its adaptability to various data resources and other application-specific requirements. For example, biases of users and items can be added in equation \ref{basic_mf}, to indicate the observed deviations of users and items, respectively. Therefore, the equation of matrix factorization with biases can be defined as:

\begin{equation}
\label{bias_mf}
\widehat{r}_{ui}=\mu + b_i + b_u + x_u^T y_i,
\end{equation}
where $\mu$ is the global average rating, and $b_u$ and $b_i$ are the deviation of user $u$ and item $i$, respectively, from the average.

At last, the biased matrix factorization approach was chosen to decompose each rating due to the following concerns. Firstly, by scrupulous investigation, it is found that the biased matrix factorization approach is superior to other recommendation algorithms such as collaborative filtering, basic matrix factorization, non-negative matrix factorization and probabilistic matrix factorization in terms of recommendation accuracy \cite{librec_tool,librec_aaai}. Secondly, complexity of the biased matrix factorization is comparable to the basic matrix factorization. In this paper, our algorithm was implemented based on the \emph{LibRec} package, which is a GPL-licensed Java library for recommender systems\footnote{http://www.librec.net/index.html} . Generally speaking, it runs much faster than other packages while achieving competitive performance. Thirdly, multiple preferences of users can be more thoroughly explored taking into account the biases of users and items. As mentioned above, users may rely on more than one aspect to make a choice, which may cause interest biases of individual users.

Given the observed ratings, the biased matrix factorization approach is learned by minimizing the objective function:
\begin{equation}
\label{bias_cost}
\min_{x^{\ast},y^{\ast},b^{\ast}} \sum_{u,i} (r_{ui} - \widehat{r}_{ui})^2 + \lambda(\| x_u\|^2 + \| y_i \|^2 + b_u^2 + b_i^2),
\end{equation}
where $\lambda$ is the regularization parameter used to prevent overfitting. In order to get the optimal parameters in equation \ref{bias_cost}, the stochastic gradient descent approach were applied to update parameters in the opposite direction of the gradient of the cost function:

% function define. NEED THIS.
%\def\ItemNN$#1${\item $\displaystyle#1$}
% update code #1
\begin{equation}
\label{update1}
\begin{aligned}
 { b }_{ u } &\leftarrow { b }_{ u }+\gamma \cdot \left( \left( { r }_{ ui }-{ \widehat { r }  }_{ ui } \right) -\lambda \cdot { b }_{ u } \right) \\
{ b }_{ i } &\leftarrow { b }_{ i }+\gamma \cdot \left( \left( { r }_{ ui }-{ \widehat { r }  }_{ ui } \right) -\lambda \cdot { b }_{ i } \right) \\
{ x }_{ u } &\leftarrow { x }_{ u }+\gamma \cdot \left( \left( { r }_{ ui }-{ \widehat { r }  }_{ ui } \right) \cdot { y }_{ i }-\lambda \cdot { x }_{ u } \right) \\
{ y }_{ i } &\leftarrow { y }_{ i }+\gamma \cdot \left( \left( { r }_{ ui }-{ \widehat { r }  }_{ ui } \right) \cdot { x }_{ u }-\lambda \cdot { y }_{ i } \right)
\end{aligned}
\end{equation}
where $\gamma$ is a step size that is usually set to a small value (e.g., 0.005).

\subsection{Rating decompositions}
Although with biased matrix factorization approach both users and items can be well modeled, its assumption that a rating is a single component would limit potential methods for the better modeling of the overall data. Therefore, we supposed that each rating is comprised of several latent components, and each component can be learned by an independent model. As mentioned above, we firstly initialized $c$ latent factor models: $\{ \Theta_1, \Theta_2, ..., \Theta_c \}$. Each model $\Theta_{\alpha}$ was based on an independent biased matrix factorization. The predictive score $\widehat{r}_{ui}^{(\alpha)}$ between user $u$ and item $i$ by model $\Theta_{\alpha}$ can be given by:

\begin{equation}
\label{sub_bias_mf}
\widehat{r}_{ui}^{(\alpha)}=\mu + b_i^{(\alpha)} + b_u^{(\alpha)} + x_u^{(\alpha)T} y_i^{(\alpha)},
\end{equation}
where $x_u^{(\alpha)}$ and $y_i^{(\alpha)}$ are latent vector of user and item with respect to model $\Theta_{\alpha}$, respectively. Secondly, cost-sensitive approach was applied to assign rating $r_{ui}$ to the latent models for training. More specifically, the following cost function was minimized as:

\begin{equation}
\label{sub_bias_cost}
\begin{aligned}
\min_{x^{\ast},y^{\ast},b^{\ast}} \sum_{u,i} \sum_{\alpha=1}^c &(w_{ui}^{(\alpha)}r_{ui}-\widehat{r}_{ui}^{(\alpha)})^2+ \\ &\lambda(\sum_{\alpha=1}^c\parallel x_u^{(\alpha)} \parallel ^2  + \sum_{\alpha=1}^c \parallel y_i^{(\alpha)} \parallel ^2 + (b_u^{(\alpha)})^2 + (b_i^{(\alpha)})^2),
 \end{aligned}
\end{equation}
where the meta-parameter $w_{ui}^{(\alpha)}$ is the weight allocated for model $\Theta_{\alpha}$ when training the rating $r_{ui}$. Meanwhile, $w_{ui}^{(\alpha)}$ was computed as follows:
\begin{equation}
\label{compute_w}
w_{ui}^{(\alpha)} = \frac{e^{-|r_{ui} - \widehat{r}_{ui}^{(\alpha)}|}}{\sum_{\alpha=1}^c e^{-|r_{ui} - \widehat{r}_{ui}^{(\alpha)}|}}.
\end{equation}

From the equation \ref{compute_w}, it can be seen that the weight of model $\Theta_{\alpha}$ is inversely proportional to the absolute error between the predictive rating $\widehat{r}_{ui}^{(\alpha)}$ and the real rating $r_{ui}$. The assumption behind our method is that the weight of model $\Theta_{\alpha}$ can be reflected by its predictive performance. Similarly, the stochastic gradient descent method was utilized to acquire parameters in equation \ref{compute_w}. In order to reduce the complexity of our method, we compute the $w_{ui}^{(\alpha)}$ by $\widehat{r}_{ui}^{(\alpha)}$ obtained in the last iterative step. The weight $w_{ui}^{(\alpha)}$ at each iteration was preserved, in order to decompose ratings further. Thus, for a given training rating $r_{ui}$, we updated the parameters in model $\Theta_{\alpha}$ as follows:

% update code #2

\begin{equation}
\label{update2}
\begin{aligned}
{ b }_{ u }^{ \left( \alpha  \right)  } &\leftarrow { b }_{ u }^{ \left( \alpha  \right)  }+\gamma \cdot \left( \left( { w }_{ ui }^{ \left( \alpha  \right)  }{ r }_{ ui }-{ \widehat { r }  }_{ ui }^{ \left( \alpha  \right)  } \right) -\lambda \cdot { b }_{ u }^{ \left( \alpha  \right)  } \right)  \\
{ b }_{ i }^{ \left( \alpha  \right)  } &\leftarrow { b }_{ i }^{ \left( \alpha  \right)  }+\gamma \cdot \left( \left( { w }_{ ui }^{ \left( \alpha  \right)  }{ r }_{ ui }-{ \widehat { r }  }_{ ui }^{ \left( \alpha  \right)  } \right) -\lambda \cdot { b }_{ i }^{ \left( \alpha  \right)  } \right)  \\
{ x }_{ u }^{ \left( \alpha  \right)  } &\leftarrow { x }_{ u }^{ \left( \alpha  \right)  }+\gamma \cdot \left( \left( { w }_{ ui }^{ \left( \alpha  \right)  }{ r }_{ ui }-{ \widehat { r }  }_{ ui }^{ \left( \alpha  \right)  } \right) \cdot { y }_{ i }^{ \left( \alpha  \right)  }-\lambda \cdot { x }_{ u }^{ \left( \alpha  \right)  } \right)  \\
{ y }_{ i }^{ \left( \alpha  \right)  } &\leftarrow { y }_{ i }^{ \left( \alpha  \right)  }+\gamma \cdot \left( \left( { w }_{ ui }^{ \left( \alpha  \right)  }{ r }_{ ui }-{ \widehat { r }  }_{ ui }^{ \left( \alpha  \right)  } \right) \cdot { x }_{ u }^{ \left( \alpha  \right)  }-\lambda \cdot y_{ i }^{ \left( \alpha  \right)  } \right)  \\
\end{aligned}
\end{equation}

Given the training rating $r_{ui}$, we re-computed the weight $w_{ui}^{(\alpha)}$ of model $\Theta_{\alpha}$ at the end of each iteration presented above. After the training process, the weight $w_{ui}^{(\alpha)}$ of model $\Theta_{\alpha}$ were then accumulated as:
\begin{equation}
\label{sum_w}
w^{\ast(\alpha)}_{ui}=\sum_{q=1}^{p}w_{ui}^{(\alpha)}(q),
\end{equation}
where $p$ is the frequency that training rating $r_{ui}$ used in the iterative process. Finally, the weight is normalized as:

\begin{equation}
\label{average_w}
\overline{w}_{ui}^{(\alpha)} = \frac{w^{\ast(\alpha)}_{ui}}{\sum_{\alpha=1}^{c}w^{\ast(\alpha)}_{ui}}.
\end{equation}
Based on the normalization weight $\overline{w}_{ui}^{(\alpha)}$, we can decompose rating $r_{ui}$ into $c$ latent components. The weight of the $\alpha th$ component can be obtained by $\overline{r}_{ui}^{(\alpha)}=\overline{w}_{ui}^{(\alpha)}r_{ui}$.

\begin{figure}[htbp]
\includegraphics[width=0.7\textwidth]{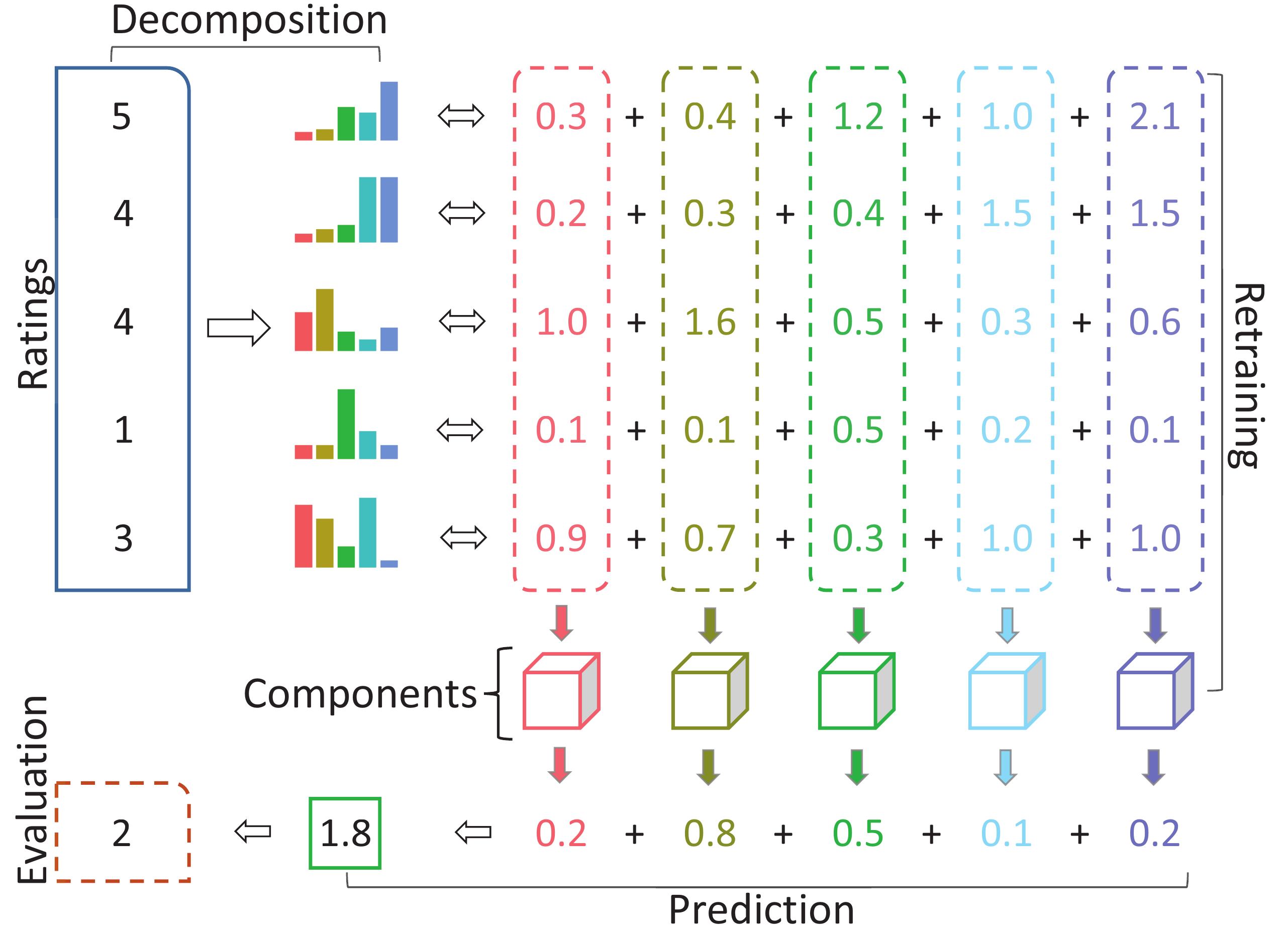}
\caption{A visualization of LCR method. The rectangle on the left side contains the original Ratings, scaled from 1 to 5. The Bar plots in the middle denote the decomposition of original ratings and the component details are shown on the right side. We use different color to denote different components of ratings and upon which, model are trained. Finally, rating in the green rectangle of the bottom is the prediction for the real rating on its left.}
\label{fig:demo}
\end{figure}

It is represented in Figure \ref{fig:demo} the visualization of LCR method, where the process of decompositions of five ratings can be seen. For example, the first rating whose value is five point, was decomposed into five latent components, namely 0.3, 0.4, 1.2, 1.0 and 2.1. Totally, five latent models were used and the average weight of these models were 0.06, 0.08, 0.24, 0.2 and 0.42, respectively, obtained by equation \ref{average_w}. Our experimental results show that the weight $w_{ui}^{(\alpha)}$ of model $\Theta_{\alpha}$ flattens out after a period of fluctuation. More details can be found in the \emph{Experiments} section.

\subsection{Models retraining}
After the above processes, $c$ learned models, as well as $c$ latent components of each rating can be obtained. For those learned models, it is different to combine them directly, since they are biased. For example, if we combine them linearly to generate predictive ratings, most of the predictive scores would exceed 5 point (supposing the rating score ranges from 1 to 5). Therefore, it is necessary to retrain those $c$ latent models. For all observed ratings, we pick out their $\alpha th$ latent component to retrain the learned model $\Theta_{\alpha}$ mentioned above. The objective function for the according model $\Theta_{\alpha}$ is similar to equation \ref{bias_cost}:

\begin{equation}
\label{retraining_cost}
\begin{aligned}
\min_{x^{\ast},y^{\ast},b^{\ast}} \sum_{u,i} &(\overline{r}_{ui}^{(\alpha)} - \widehat{r}_{ui}^{(\alpha)})^2 + \\ &\lambda(\| x_u^{(\alpha)} \|^2 + \| y_i^{(\alpha)} \|^2 + (b_u^{(\alpha)})^2 + (b_i^{(\alpha)})^2),
\end{aligned}
\end{equation}
where $r_{ui}$ in equation \ref{bias_cost} is replaced by latent component $\overline{r}_{ui}^{(\alpha)}$. We represented this process for each learned model. It is worth mentioning that the learned model $\Theta_{\alpha}$ can't be re-initialized. We tried to re-initialized the learned models, but failed. The model retraining process can also be found in Figure \ref{fig:demo}. For instance, the first latent component of each rating (marked as the red color in the first column) was selected to retrain the latent model $\Theta_1$. Since we assumed that those latent models were independent to each other, other latent components were not used when the model $\Theta_1$ was retrained.

\subsection{Rating predictions}
Then, $c$ new latent factor models were obtained. By combining those models linearly, we can get the final predictive score between user $u$ and item $i$:
\begin{equation}
\widehat{r}_{ui}=\sum_{\alpha=1}^c \widehat{r}_{ui}^{(\alpha)},
\end{equation}
where $\widehat{r}_{ui}^{(\alpha)}$ is computed by equation \ref{sub_bias_mf}.

In general, our method can be divided into two steps: rating decomposition and model retraining. Our method is more complicated than the tradition matrix factorization method, but its running time grows linearly with the traditional MF. The details can be found in the \emph{Experiments} section. The visual representation of our method is presented in Figure \ref{fig:demo}.

\section{Experiments}
\label{Exp}
\subsection{Experimental Setup}
In order to evaluate the accuracy of our method, five benchmark datasets are selected, namely MovieLens, Douban, Movietweetings, Epinions and Goodreads. MovieLens is a movie recommendation website, which uses individual users' rating to generate personalized recommendations \cite{dataset_MovieLens}. Douban, launched on March 6, 2005, is a Chinese Web 2.0 web site providing user reviews and recommendation services of movies, books, and music \cite{Huang_ESI_2012}. The raw data contains user activities before Aug 2010 and we reserve those users who have rated more than $20$ movies. Movietweetings is a dataset consisting of ratings on movies that were contained in well-structured tweets on Twitter \cite{dataset_Movietweetings}. The raw dataset consists of $45,604$ users and $26,165$ items. Epinions is a web site where people can review products and the raw data consists of $876,252$ users and $120,492$ items. Goodreads is a book sharing and recommendation website and the collected data contains user activities before we crawled the website on Aug 2010. As it is quite difficult to provide accurate recommendations for inactive users, we filtered out users who have rated less than 20 items. The detailed information of datasets are presented in Table \ref{tab:data}.

\setlength{\textfloatsep}{10pt plus 1.0pt minus 2.0pt}
\begin{table}[htbp]
\def\arraystretch{1.6}
\begin{tabular}{l|c|c|c}
\hline
Dataset & \#Users,$n$ & \#Items,$m$ & \#Ratings,$l$    \\ \hline \hline
MovieLens & 6,040 & 3,706 & 1,000,209 \\
Douban & 6,472 & 7,755 & 2,147,843 \\
Movietweetings & 2,331 & 1,669 & 196,359 \\
Epinions & 66,512 & 12,631 & 5,909,085 \\
GoodReads & 96,131 & 39,704 & 12,577,677\\ \hline
\end{tabular}
\caption{The statistics of five benchmark datasets, MovieLens, Douban, Movietweetings, Epinions and GoodReads. The table shows the number of user($n$), number of items($m$) and number of ratings($l$) for each dataset accordingly.}
\label{tab:data}
\end{table}

Each dataset is randomly divided into two parts: the training set ($E^T$) and the probe set ($E^P$). The training set and the testing set are chosen via cross-validation \cite{train_test_split}. The commonly used Root Mean Square Error (RMSE) was adopted to evaluate the accuracy of methods, which could be expressed as
\begin{equation}
\label{eq:rmse}
{ RMSE }=\sqrt { \frac { 1 }{ \left| E^P \right|  } \sum _{ { r }_{ ui }\in E^P }^{  }{ { \left( { \hat { r }  }_{ ui }-{ r }_{ ui } \right)  }^{ 2 } }  } .
\end{equation}

\subsection{Comparison Models}
Eight methods in total were selected to compare with our method. \emph{Slopone} is an item-based collaborative filtering approach which was chosen as the benchmark method \cite{Exp_SlopOne}. Five matrix factorization methods were selected, which were \emph{biased matrix factorization} (BMF for short), \emph{SVD++}, \emph{non-negative matrix factorization} (NMF for short), \emph{probabilistic matrix factorization} (PMF for short) and \emph{Bayesian probabilistic matrix factorization} (BPMF for short). Two cluster-based methods, which were \emph{Latent Dirichlet Bayesian co-clustering method} (LDCC for short) and \emph{Bayesian user community model} (BUCM for short). The cluster-based method partitions individual ratings in a macro-level, whereas our method decomposes ratings in a micro-level. Therefore, we adopted those two cluster-based methods to compare with our method. Details of these methods are given as follows:

\begin{itemize}%for capital roman numbers.
\item \textbf{SlopOne}:  A simple and strong memory based model that separately factors in items which user liked and disliked \cite{Exp_SlopOne}. SlopeOne is the simplest form of non-trival item-based collabortive filtering algorithm so far, which is the benchmark.
\item \textbf{Biased Matrix Factorization (BMF)}: A latent factor model that directly models only the oberserved ratings and prevents overfitting with a regularized term \cite{Exp_biasedMF}. BMF is adopted as the base model of LCR. Since our method has more parameters than BMF, we train multi BMF models and average their prediction results (MBMF for short). This method is chosen on condition that it shares the same number of parameters with our method.
\item \textbf{SVD++}: A latent factor model which makes use of implicit feedback information of users \cite{Exp_biasedMF}. Generally speaking, implicit feedback refers to users' history information which indicate their preference.
\item \textbf{Non-negative matrix factorization (NMF)}: A matrix factorization method that constraints its factorization results to be nonnegative. The learned nonnegative vectors are sparse representation of the users and items \cite{Exp_NMF}.
\item \textbf{Probabilistic Matrix Factorization (PMF)} and \textbf{Bayesian Probabilistic Matrix Factorization (BPMF)}: As a probabilistic view of matrix factorization, the PMF is able to scale linearly with the observed ratings and has good performance on very sparse and imbalance data \cite{PMF}. The BPMF can prevent overfitting through integrating over all model parameters and hyperparameters \cite{Bayesia_PMF}.
\item \textbf{Latent Dirichlet Bayesian Co-Clustering (LDCC)}: A co-clustering methods allows users and items to be grouped simultaneously. The LDCC matches ratings in terms of users and items \cite{Exp_LDCC}.
\item \textbf{Bayesian User Community Model (BUCM)}: Relied on both item selection and rating emission, the BUCM generate communities for users that experience the same items and individual user  is modeled as mixture of topics which characterized by both items and communities for users \cite{Exp_BUCM}.
\end{itemize}

\subsection{Results and analysis}

\begin{figure}[ht]
\includegraphics[width=1.0\textwidth, height=0.4\textwidth]{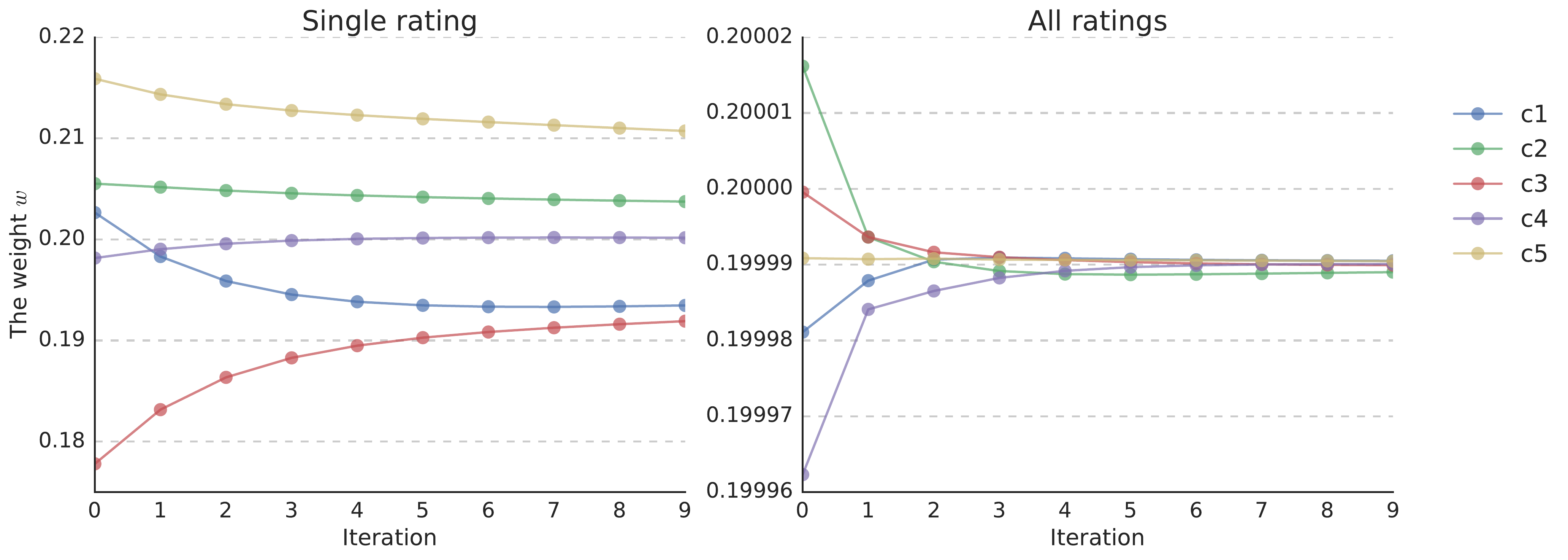}
\caption{The convergence of $w$, i.e., the weights of rating components. The left plot shows the convergence of $w$ in a single rating entity and the right plot shows that with respect to all ratings, that is, the average of all entities. The weight for different components are denoted by different colors.}
\label{fig:weight_convergence}
\end{figure}

\textbf{The convergence of weights of models.} In our method, the weight $w_{ui}^{\alpha}$ obtained by equation \ref{compute_w} is used to decompose a rating. Therefore, the stability of our method would be significantly affected by the convergency of $w_{ui}^{\alpha}$. Thus we randomly selected a rating and plotted the convergency of $w_{ui}^{\alpha}$ as shown in the left of Figure \ref{fig:weight_convergence}. In total, five latent factor models were exploited to decompose ratings. From Figure \ref{fig:weight_convergence}, it can be seen that the weights of models stay stable after the third iteration. In addition, it can be indicated by the varied weights of latent models that users may have complex preferences. Moreover, the average weight of each latent model over all ratings was computed and the results can be found in the right of Figure \ref{fig:weight_convergence}. It can be seen that there are almost no difference among the average weights of latent models. In other words, in a micro level (a single rating), the latent models are different from each other, but in a macro level (all ratings), they are unbiased.

\begin{figure}[ht]
\includegraphics[width=0.95\textwidth]{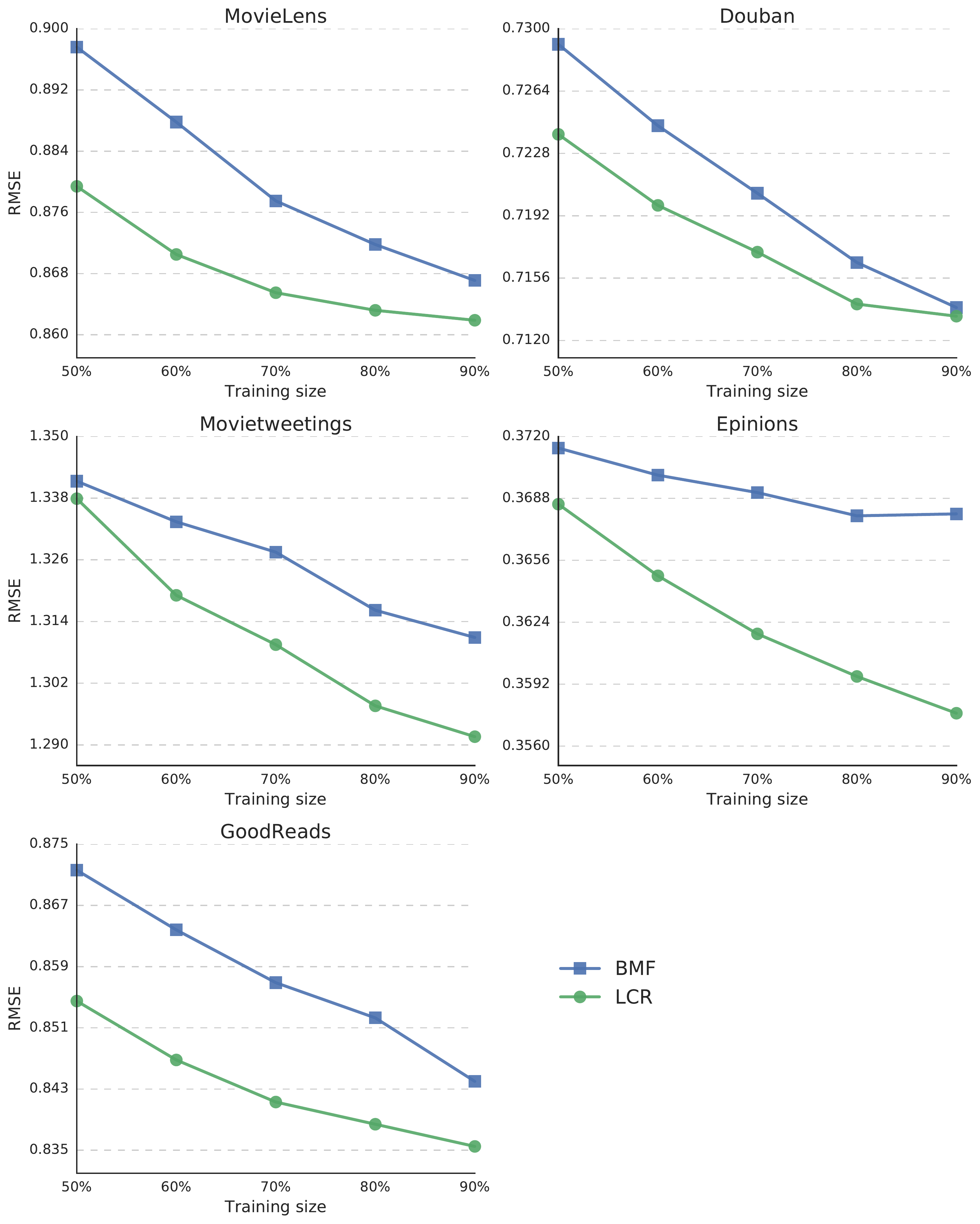}
\caption{The relationship between LCR's accuracy and the sparsity of dataset are shown in the plots. Each plot represents the prediction results in one dataset on RMSE. We use blue-square line and green-circle line to denote BMF and LCR respectively.}
\label{fig:lcr2}
\end{figure}

\begin{figure}
\includegraphics[width=0.6\textwidth, height=0.45\textwidth]{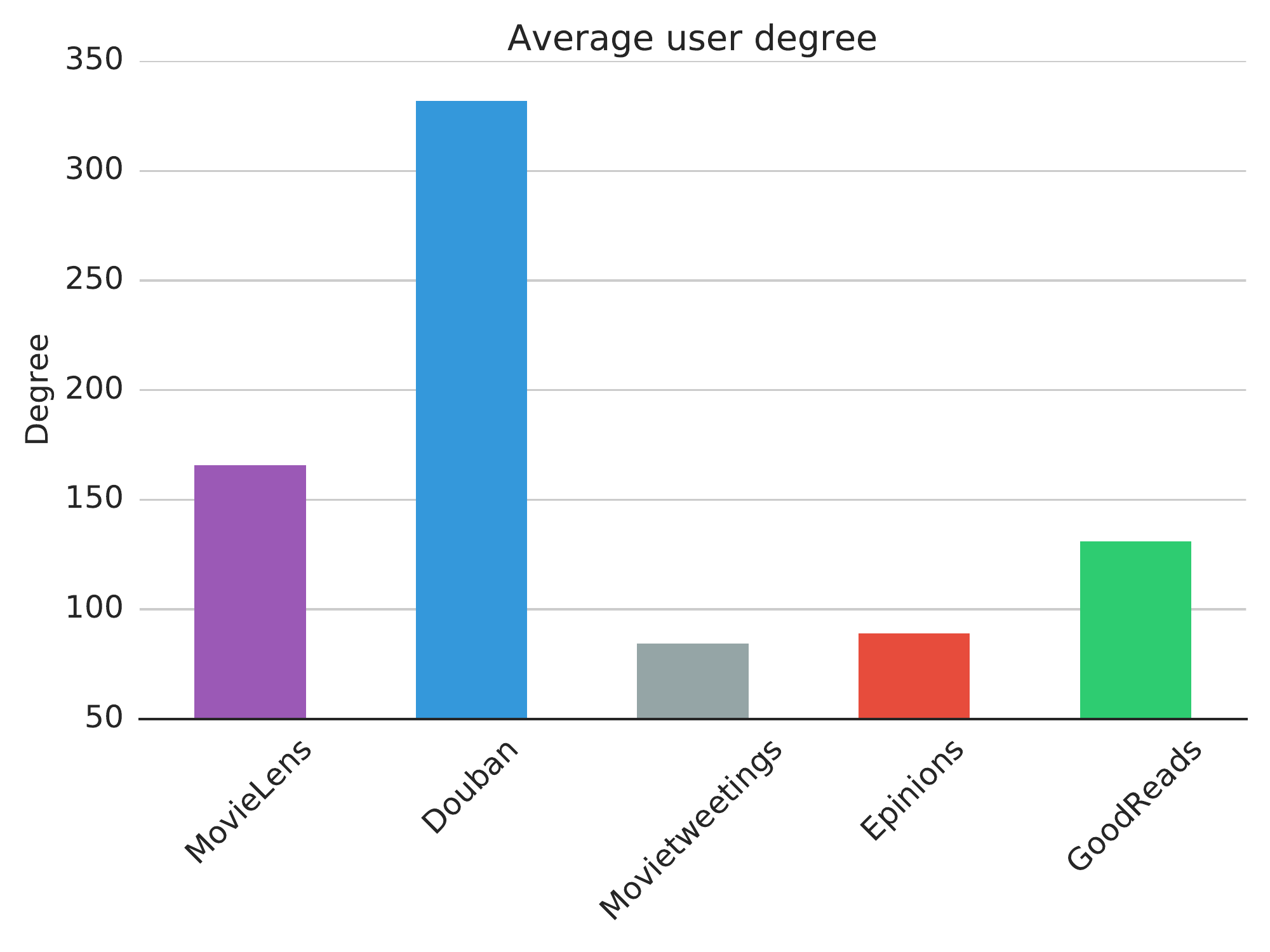}
\caption{The average degree of users for all datasets. It can be seen that the average degree of users in Movietweeting and Epinions are way smaller than the other datasets.}
\label{fig:network_property}
\end{figure}

\textbf{The performance of LCR.} Meanwhile, we compared our method with the-state-of-the-art approaches. The results are presented in Table \ref{tab:accuracy}, where it can be seen that our method enjoys the best predictive accuracy in all datasets. It's worth mentioning that the benchmark method SlopOne outperforms some matrix factorization based methods in Douban and Epinions datasets. It may be caused by the sparsity of datasets \cite{Cacheda_CCF_2011}. In other words, the matrix factorization based methods require adequate data for training to generate accurate recommendations. Nevertheless, our method is better than SlopOne in both relatively dense and sparse datasets, which indicates that our method can overcome the sparsity problem of recommender systems to some extent. Motivated by the results in Table \ref{tab:accuracy}, we explored the relationship between the accuracy of our method and the sparsity of datasets further. The results are presented in the bottom of Figure \ref{fig:lcr2}, in which the x-axis denotes the ratio of the training set to the whole dataset (training size) and the y-axis is the accuracy of algorithms running on the corresponding training set. For MovieLens, Douban and Goodreads datasets, the improvement of our method is more significant when taking into account fewer ratings. However, in Epinions and MovieTweeting datasets, our method is more accurate than the biased matrix factorization method when considering more ratings.
In order to find the possible reason, we studied the user degree which is defined as the number of items that the user has collected, as presented in Figure \ref{fig:network_property}. It can be seen that the average user degree of Epinions and MovieTweeting is smaller than that in other datasets, which means users of Epinions and MovieTweeting usually collect less items than those in other datasets. Therefore, it requires more ratings to uncover users' complex preferences for these two datasets. In addition, from Table \ref{tab:accuracy} one can see that our method outperforms MBMF which shares the same number of parameters with LCR. This result indicates that improvements of our method are not caused by adding more parameters.

% \begin{figure}[ht]
% \includegraphics[width=18cm,height=7cm]{figure3_LCR2.pdf}
% \caption{The relationship between LCR's accuracy and the number of latent components are presented in the top and the relationship between LCR's accuracy and the sparsity of dataset are shown in the bottom.}
% \label{fig:lcr2}
% \end{figure}

\begin{figure}[ht]
\includegraphics[width=0.95\textwidth]{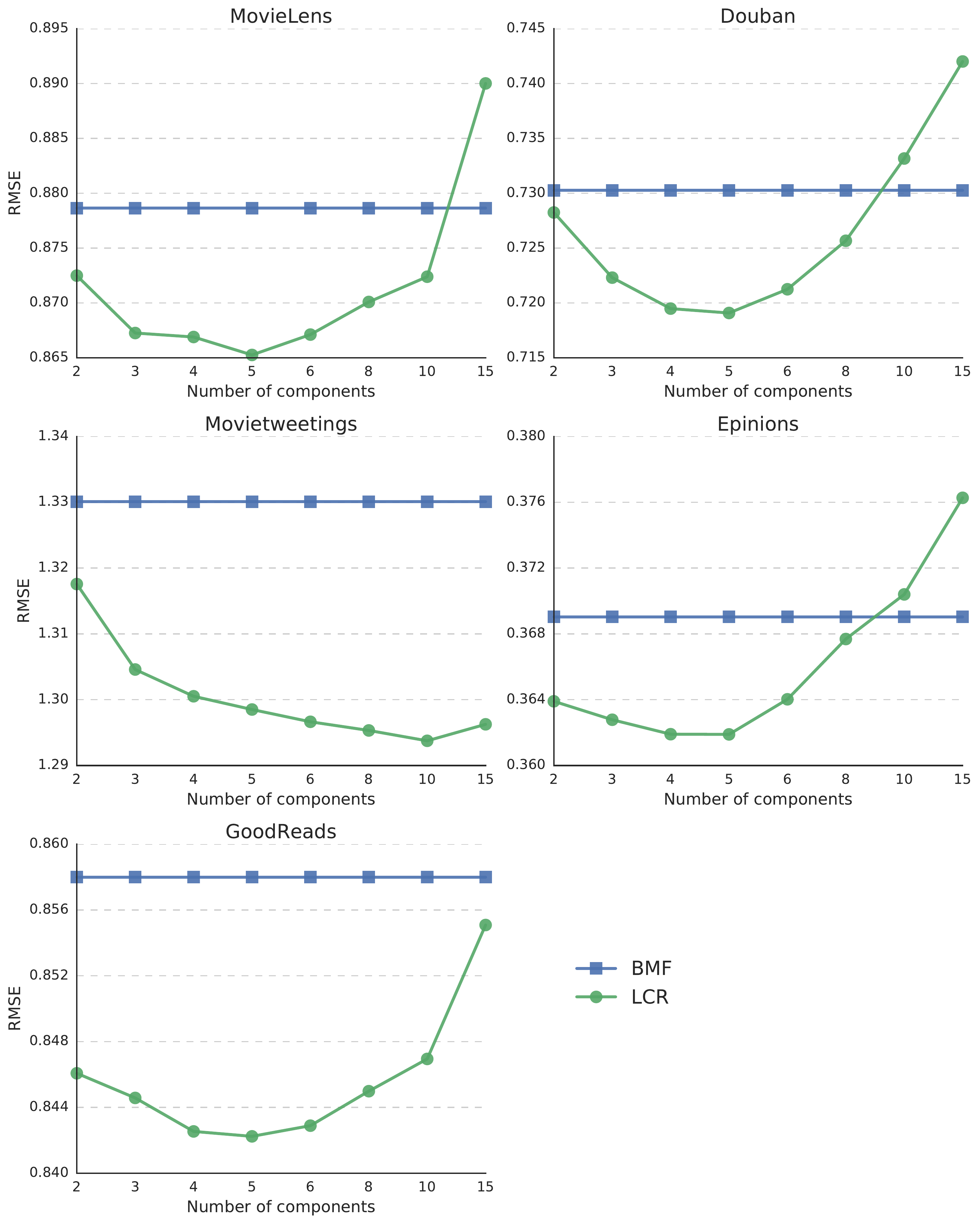}
\caption{The relationship between LCR's accuracy and the number of latent components are presented in the plots. Each plot represents the prediction results in one dataset on RMSE. We use blue-square line and green-circle line to denote BMF and LCR respectively.}
\label{fig:lcr1}
\end{figure}

\begin{table}
\def\arraystretch{1.6}
\begin{tabular}{lp{1.7cm}p{1.7cm}p{1.7cm}p{1.7cm}p{1.7cm}}
\hline
 & MovieLens & Douban & Movietweetings & Epinions & GoodReads  \\ \hline \hline
SlopOne & 0.9024(.0013) & 0.7266(.0005) & 1.3266(.0043) & 0.3627(.0008) & 0.8602(.0003) \\
BMF & 0.8786(.0014) & 0.7303(.0003) & 1.3301(.0073) & 0.3691(.0006) & 0.8580(.0002) \\
MBMF & 0.8774(.0019) & 0.7303(.0003) & 1.3002(.0065) & 0.3690(.0006) & 0.8567(.0005)\\
NMF & 0.9322(.0026) & 0.7458(.0065) & 1.4067(.0069) & 0.4071(.0005) & 0.8959(.0004) \\
PMF & 0.8875(.0018) & 0.7371(.0012) & 1.4312(.0032) & 0.3908(.0029) & 0.8737(.0016) \\
BPMF & 0.8786(.0015) & 0.7227(.0002) & 1.4064(.0032) & 0.3752(.0009) & 0.8590(.0002) \\
SVD++ & 0.8729(.0021) & 0.7266(.0005) & 1.3211(.0059) & 0.3672(.0007) & 0.8535(.0006) \\
LDCC & 0.9387(.0066) & 0.7472(.0009) & 1.4413(.0040) & 0.3820(.0019) & 0.8883(.0003) \\
BUCM & 0.9605(.0057) & 0.7775(.0008) & 1.4711(.0295) & 0.3910(.0116) & 0.9404(.0010) \\
LCR & \contour{black}{0.8653(.0008)} & \contour{black}{0.7191(.0001)} & \contour{black}{1.2937(.0005)} & \contour{black}{0.3619(.0001)} & \contour{black}{0.8422(.0005)} \\ \hline
\end{tabular}
\caption{The accuracy of recommendation methods with respect to \textit{RMSE} with the standard error shown within the brackets.}
\label{tab:accuracy}
\end{table}

\textbf{The number of the latent components.} In our method, the number of latent component is in fact a parameter, which can affect the performance of the method if a proper value is not given. Thus, we accordingly investigated the relationship between the accuracy of our method and the number of latent components of ratings, as shown in the top of Figure \ref{fig:lcr1}. Generally speaking, the accuracy of algorithm can be enhanced when ratings are decomposed into more than two components. The results verified our former hypothesis that users' interests are complex and multi-dimensional. Furthermore, it can be seen that the optimal number of latent components is around 5 (except Movietweeting data which is around 10), which indicates that our method is not sensitive to the number of latent components.

\textbf{The complexity of LCR.}
As mentioned above, the running time of our method grows linearly with the traditional matrix factorization. It can be found in Table \ref{tab:time} the comparison between the running time of LCR and its based model BMF. As it is mentioned in section 3, the LCR model requires rating decomposition and retraining to accomplish its lifecycle. In rating decomposition, LCR requires a minimization process by equation \ref{sub_bias_cost} to learn the model weights of each component, which consumes one BMF time. In retraining, LCR needs to retraining all desired latent component models with equation \ref{retraining_cost}. For each latent component model, LCR needs one BMF time. It turns out that the running time of LCR is approximately $1+C$ times to the standard matrix factorization model, where $C$ is the number of desired latent components. The $1+C$ times ratio can be reflected by our running time experiments. In Table \ref{tab:time}, taking LCR$^5$(with 5 latent components) for example, the running time of LCR is approximately six to seven times to that of BMF in all datasets. In practice, the running time ratio might exceed a bit due to the normalization for the weight of components during the operation of algorithms in equation \ref{compute_w}, \ref{sum_w}, \ref{average_w}. Thus, it is verified that the running time of our model grows linearly with the traditional matrix factorization. Although our method brings additional computation cost comparing to the single model, our method has a stable performance even in a relatively sparse data set (e.g. Douban and Movietweetings). That is, combination of several models would offset the bias caused by the sparsity of data set.

\begin{table}[!htbp]
\def\arraystretch{1.6}
\begin{tabular}{{p{1.85cm}p{0.75cm}}*{4}{p{0.80cm}}}
\hline
& BMF & LCR$^2$ & LCR$^3$ & LCR$^4$ & LCR$^5$ \\ \hline \hline
MovieLens 	& 24.0 & 80.6 & 111.4 & 143.3 & 184.8 \\
Douban 		& 48.9 & 173.5 & 234.2 & 327.4 & 400.9 \\
Movietweetings & 6.9 & 24.5 & 31.6 & 40.5 & 50.3 \\
Epinions 		& 167.6 & 576.8 & 813.7 & 1131.1 & 1327.6 \\
GoodReads  	& 574.2 & 1717.6 & 2387.8 & 3003.5 & 3482.2 \\
\hline
\end{tabular}
\caption{The runing time of LCR, in seconds. We compare the running time of LCR to its base model BMF. The number of components for LCR is denoted by the superscript, for e.g. LCR$^3$ denoted LCR with three components.}
\label{tab:time}
\end{table}

\section{Conclusion}
\label{Conclusion}
In this paper, a user is assumed to have multi-dimensional preferences and whose ratings are consist of multiple independent latent components. To determine the latent components, a new recommender system, called LCR, was proposed to learn the user preferences by exploiting the latent components of user ratings. Based on the cost-sensitive learning strategy, LCR is capable of decomposing an existing user rating into several components. Meanwhile, each latent component is treated as an independent part, with which LCR can better model user preferences at an individual level and distinguish itself from existing methods. Moreover, it is verified by extensive experiments and analysis on several benchmark datasets that our model can (1) make more accurate recommendations by introducing proper latent rating components, (2) perform well even provided with fewer ratings and (3) generate item recommendations at the cost of a linearly increased running time comparing to the single model.
\begin{acknowledgements}
This work was supported by the National Natural Science Foundation of China (61502078) and Scientific Research start-up Foundation (ZYGX2015KYQD073).
\end{acknowledgements}

\bibliographystyle{spmpsci}      % mathematics and physical sciences
\bibliography{LCR_proc.bib}   % name your BibTeX data base

%\begin{thebibliography}{}

%\bibitem{handbook}
%Francesco Ricci, Lior Rokach,Bracha Shapira, Article title, Journal, Volume, page numbers (year)
% etc
%\end{thebibliography}

\end{document}